\documentclass[aps,prb, twocolumn, superscriptaddress]{revtex4}
\usepackage{amsmath,amsfonts, amsthm, amssymb}
\usepackage{color}
\usepackage{bm}
\usepackage{graphicx, subfigure}
\usepackage{subfigure}
\usepackage{verbatim}
\usepackage{mathrsfs}
\usepackage{hyperref}
\usepackage{dsfont}
\usepackage{pst-all}
\usepackage{tikz}
\usetikzlibrary{calc}

\begin{document}

\newcommand{\odiff}[2]{\frac{\di #1}{\di #2}}
\newcommand{\pdiff}[2]{\frac{\partial #1}{\partial #2}}
\newcommand{\di}{\mathrm{d}}
\newcommand{\ii}{\mathrm{i}}
\renewcommand{\vec}[1]{{\mathbf #1}}
\newcommand{\vx}{{\bm x}}
\newcommand{\ket}[1]{|#1\rangle}
\newcommand{\bra}[1]{\langle#1|}
\newcommand{\pd}[2]{\langle#1|#2\rangle}
\newcommand{\tpd}[3]{\langle#1|#2|#3\rangle}
\renewcommand{\vr}{{\vec{r}}}
\newcommand{\vk}{{\vec{k}}}
\renewcommand{\ol}[1]{\overline{#1}}
\newtheorem{theorem}{Theorem}
\newcommand{\comments}[1]{ }
\newcommand{\mb}[1]{\mathbf{#1}}
\newcommand{\newsection}[1]{\section{#1}}
\newcommand{\Ref}[1]{Ref. [\onlinecite{#1}]}
\newcommand{\Refs}[1]{Refs. [\onlinecite{#1}]}
\newcommand{\ntxt}[1]{\textcolor{blue}{#1}}

\title{Classification of Symmetry-Protected Phases for Interacting Fermions in Two Dimensions}
\author{Meng Cheng}
\affiliation{Department of Physics, Yale University, New Haven, CT 06511-8499, USA}
\affiliation{Station Q, Microsoft Research, Santa Barbara, CA 93106, USA}
\author{Zhen Bi}
\affiliation{Department of Physics, Massachusetts Institute of Technology, Cambridge, MA 02139, USA}
\affiliation{Department of Physics, University of California, Santa Barbara, CA 93106, USA}
\author{Yi-Zhuang You}
\affiliation{Department of Physics, Harvard University, Boston, MA 93106, USA}
\affiliation{Department of Physics, University of California, Santa Barbara, CA 93106, USA}
\author{Zheng-Cheng Gu}
\affiliation{Department of Physics, The Chinese University of Hong Kong, Shatin, New Territories, Hong Kong}
\affiliation{Perimeter Institute for Theoretical Physics, Waterloo, ON, N2L 2Y5, Canada}
\date{\today}

\begin{abstract}
	Recently it has been established that two-dimensional bosonic symmetry-protected topological(SPT) phases with on-site unitary symmetry $G$ can be completely classified by the group cohomology $H^3(G, \mathrm{U}(1))$. Later, group super-cohomology was proposed as a partial classification for SPT phases of interacting fermions. In this work, we revisit this problem based on the algebraic theory of symmetry and defects in two-dimensional topological phases.  We reproduce the partial classifications given by group super-cohomology, and we also show that with an additional $H^1(G, \mathbb{Z}_2)$ structure, a complete classification of SPT phases for two-dimensional interacting fermion systems with a total symmetry group $G\times\mathbb{Z}_2^f$ is obtained. We also discuss the classification of interacting fermionic SPT phases protected by time reversal symmetry.
\end{abstract}

\maketitle

\newsection{Introduction}
Due to the Pauli exclusion principle, non-interacting fermions have rich structures in their ground state wavefunctions and a complete classification of symmetry protected topological(SPT) phases of free fermions has been achieved by using ideas like Anderson localization~\cite{Schnyder2008} and K theory~\cite{Kitaev2009}.  
A variety of materials have been experimentally discovered to realize these nontrivial SPT phases, such as time-reversal-invariant topological insulators/superconductors~\cite{3dTI, QSH2007, Hasan_RMP2010, Qi_RMP2011}.
On the other hand, recent studies show that SPT phases also exist in interacting boson systems and can be systematically classified by (generalized) group cohomology theory~\cite{Chen_science, Chen_arxiv2011, Wen_arxiv2014}. Strong interaction is a necessity for the occurrence of bosonic SPT phases.

Despite the remarkably successful classifications of non-interacting Hamiltonians, the non-perturbative effects of interactions in fermionic SPTs still remain an important theoretical question. In many cases, the free-fermion classifications are shown to be stable against interaction effects, e.g. the $\mathbb{Z}_2$ classification of time-reversal-invariant topological insulators~\cite{RyuPRB2012}. A breakthrough by Fidkowski and Kitaev~\cite{Fidkowski_PRB2011} demonstrated that in one-dimensional fermionic systems with time-reversal symmetry $T^2=1$, the non-interacting $\mathbb{Z}$ classification breaks down to $\mathbb{Z}_8$ when strong interactions are present~\cite{Fidkowski_PRB2011, Turner_PRB2011}. The result was then generalized to two dimensions with an on-site $\mathbb{Z}_2$ symmetry~\cite{Qi_arxiv2012, Ryu_arxiv2012, Yao_arxiv2012,Lu_arxiv2012, Gu_2013, NeupertPRB2014} and three-dimensional time-reversal-invariant topological superconductors~\cite{KitaevZ16, Metlitski_arxiv2014,Wang_PRB2014, Chen_PRB2014, Fidkowski_PRX2013, YouXu}. Recently several general classification schemes have been proposed for fermionic SPT phases, including group super-cohomology~\cite{GuSuperCoho, GK}, spin cobordism~\cite{KapustinFSPT} and invertible topological field theories~\cite{Freed2014, Freed2}.

In this work we pursue an alternative route to classify fermionic SPT(fSPT) phases in 2D. Following previous works on bosonic SPT phases, to characterize fSPT phases we introduce \textit{extrinsic} defects carrying symmetry fluxes into the fSPT state. The classification is obtained by studying the topological properties of the defects, such as their fusion rules and braiding statistics.  Similar ideas have proven to be quite successful in classifying bosonic SPT phases in 2D~\cite{LevinGu_arxiv2012, ChengPRL2014, BBCW, HungPRB2014}. The mathematical objects that classify 2D fSPT phases can be summarized as three group cohomologies of the symmetry group: $H^1(G,\mathbb{Z}_2)$, $BH^2(G,\mathbb{Z}_2)$ and $H^3(G,\mathrm{U}(1))$. $H^1(G,\mathbb{Z}_2)$ classifies fSPT phases with unpaired Majorana edge modes.  $BH^2(G, \mathbb{Z}_2)$ was previously derived in the group super-cohomology classification~\cite{GuSuperCoho}, and we clarify its physical meaning as the fractionalized symmetry quantum numbers carried by fermion-parity $\pi$ fluxes. Finally, $H^3(G,\mathrm{U}(1))$ is the known classification of bosonic SPT phases~\cite{Chen_science}. {To gain more physical intuition for the above results, let us consider the simplest case with $G=\mathbb{Z}_2$ which has a $\mathbb{Z}_8$\cite{Gu_2013,KapustinFSPT,Freed2014} classification. In this case, $H^3(\mathbb{Z}_2,\mathrm{U}(1))=\mathbb{Z}_2$ gives the classification of bosonic SPT phases when all fermions form bosonic moleculars/spins; $BH^2(\mathbb{Z}_2,\mathbb{Z}_2)=\mathbb{Z}_2$ classifies the fractionalized $\mathbb{Z}_2$ quantum numbers carried by fermion-parity $\pi$ fluxes; and finally $H^1(\mathbb{Z}_2,\mathbb{Z}_2)=\mathbb{Z}_2$ indicates whether the $\mathbb{Z}_2$ symmetry fluxes carry unpaired Majorana zero modes or not. All together, there are $8$ fSPT phases and further analysis suggests a $\mathbb{Z}_8$ group structure.}

\newsection{Generalities} First of all, let us clarify the meaning of the symmetry group $G$ in a fermionic system. A fundamental symmetry of fermionic systems that can never be broken is the conservation of total fermion parity, denoted by $\mathbb{Z}_2^f$. In addition to this symmetry, we assume the system has an on-site symmetry group $G$. The total symmetry group of the system is actually $\mathbb{Z}_2^f\times G$~\footnote{In this definition we exclude symmetries such as the $\mathrm{U}(1)$ symmetry of fermion number conservation, of which $\mathbb{Z}_2^f$ is a subgroup.}. For most of our paper we assume $G$ is unitary and finite. We will consider anti-unitary time-reversal symmetry in the end.

Our approach to the problem is based on the algebraic theory of two-dimensional gapped quantum phases~\cite{Kitaev2006}.  Given a gapped phase in two dimensions, one can classify the low-energy localized quasiparticle excitations into superselection sectors (topological charges). Different topological charges can not be transformed into each other by applying local operators. Fusion and braiding of the topological charges are described by the mathematical framework of unitary braided tensor category (UBTC)~\cite{Kitaev2006, Bonderson07}.
From this point of view, a bosonic SPT phase has no topologically nontrivial quasiparticle excitations, and the corresponding UMTC is just the trivial one: $\mathcal{C}=\{I\}$, where $I$ stands for the vaccum sector.
In a fermionic system, states with even and odd numbers of fermions belong to different superselection sectors, and no terms in the Hamiltonian can ever mix these two sectors. Therefore we model a gapped fermion SPT phase abstractly by the (premodular) UBTC $\mathcal{C}_f=\{I,\psi\}$, where $I$ stands for trivial bosonic excitations (sometimes refered to as the vaccum) and $\psi$ represents a single fermionic excitation. $\psi$ still represents a local excitation, but distinct from the (bosonic) vaccum. The fusion rule is obviously $\psi\times\psi=I$. 

To classify fSPT phases with a unitary symmetry $G$, we exploit the idea of gauging the symmetry ~\cite{LevinGu_arxiv2012, Gu_2013, ChengPRL2014} and use topological properties of the extrinsic point-like defects carrying symmetry fluxes to distinguish different fSPT phases.  The defining feature of symmetry defects is a generalized Aharonov-Bohm effect: when a quasiparticle excitation is transported around a $\mathbf{g}$-defect with $\mathbf{g}\in G$, it must be transformed by the local symmetry operation corresponding to $\mathbf{g}$. They can be introduced into the system by explicitly modifying the Hamiltonian along the defect branch cut,  see [\onlinecite{BBCW}].  We then enlarge our algebraic theory of quasiparticle excitations to include these defects. Defects by construction carry group labels, but they can also have their own topological charge labels, i.e. how charge types are permuted by symmetry transformations. We therefore collect all defects labeled by the same group element $\mathbf{g}\in G$ into a $\mathbf{g}$-sector $\mathcal{C}_\mathbf{g}$, and define the so-called $G$-extension $\mathcal{C}_G$ as $\mathcal{C}_G=\bigoplus_{\mb{g}\in G} \mathcal{C}_\mb{g}$.
Notice that the $\mb{g}=1$ sector $\mathcal{C}_1$ is just the original theory $\mathcal{C}_f$. Similar to anyon models, the most fundamental property of defects is their fusion rules, i.e. how the topological charges are combined. But because defects also carry group labels, their fusion rules must respect the group multiplication structure (i.e. $G$-graded), namely for $a_\mathbf{g}\in \mathcal{C}_\mb{g}, b_\mb{h}\in\mathcal{C}_\mb{h}$, we have
 \begin{equation}
	 a_\mathbf{g}\times b_\mb{h}=\sum_{c_\mb{gh}}N_{a_\mb{g}b_\mb{h}}^{c_\mb{gh}}c_\mb{gh},
	 \label{}
 \end{equation}
 where $N_{a_\mb{g}b_\mb{h}}^{c_\mb{gh}}$ are non-negative integers indicating the number of ways defects $a_\mb{g}$ and $b_\mb{h}$ can combine to produce charge $c_\mb{gh}$.

 In order to completely define the $G$-extension, we need to study more subtle structures, such as the associativity of defect fusion and braiding transformations of defects. These are captured in a consistent mathematical formalism called $G$-crossed braided fusion category, and we refer the readers to Ref. [\onlinecite{BBCW}] for a thorough discussion. In our case, we follow a more physically intuitive argument to avoid solving complicated algebraic equations.
 In particular, we will make use of the ``invertibility'' property of SPT phases: for each SPT state there is a \textit{unique} ``conjugate'' state such that by stacking them up one obtains the trivial state~\cite{Freed2014}.  Furthermore, given two SPT states, one can stack them together to get another SPT state, which is defined as their sum. More precisely,  given two SPT phases described by Hamiltonians $H_1$ and $H_2$, their sum is defined as the ground state of $H_1\oplus H_2$. For $\mb{g}\in G$, we define the enlarged symmetry operation $U(\mb{g})=U_1(\mb{g})\otimes U_2(\mb{g})$ where $U_{1,2}(\mb{g})$ are the corresponding symmetry operations in the subsystems. In other words, SPT phases can be naturally endowed with an Abelian group structure. We will denote the Abelian group of fSPT phases with a given symmetry group $G$ by $\mathscr{G}$.

\newsection{Classifying defect fusion rules}
The $G$-grading structure of the fusion rules of defects has a profound consequence : one can show that all sectors $\mathcal{C}_\mb{g}$ have the same total quantum dimensions: $\mathcal{D}_\mb{g}^2=\mathcal{D}_1^2=2$ where $\mathcal{D}_\mb{g}^2=\sum_{a_\mb{g}\in \mathcal{C}_\mb{g}}d_{a_\mb{g}}^2$~\cite{ENO, BBCW}. In our case, $\mathcal{D}_\mb{g}^2=2$ leaves us with only two options: (a) There are two Abelian defects in $\mathcal{C}_\mb{g}$ and they differ by fusing with $\psi$. We denote them by $\sigma_{\mb{g}}^\pm$. (b) There is a single non-Abelian defect in $\mathcal{C}_\mb{g}$ with quantum dimension $\sqrt{2}$. We denote it by $\sigma_\mb{g}$.

First we show that the possible non-Abelian fusion rules have one-to-one correspondence with $H^1(G,\mathbb{Z}_2)$, i.e. group homomorphisms from $G$ to $\mathbb{Z}_2$.
Assume for both $\mb{g,h}\in G$ the defects are non-Abelian. To be able to construct the fusion outcome of $\sigma_\mb{g}\times\sigma_\mb{h}\in \mathcal{C}_{\mb{gh}}$, we immediately see that the defects in the $\mb{gh}$ sector must be Abelian just to match the quantum dimension. There are still three possibilities: $\sigma_\mb{g}\times\sigma_\mb{h}=2\sigma_{\mb{gh}}^+,\sigma_\mb{g}\times\sigma_\mb{h}=2\sigma_{\mb{gh}}^-$ and $\sigma_\mb{g}\times\sigma_\mb{h}=\sigma_{\mb{gh}}^+ +\sigma_{\mb{gh}}^-$. The former two are impossible for the following reason: Assuming $\sigma_\mb{g}\times\sigma_\mb{h}=2\sigma_{\mb{gh}}^+$. Using the symmetry of the fusion coefficients~\footnote{Generally we have $N_{ab}^c=N_{\ol{a}b}^c=N_{c\ol{b}}^a$~\cite{Kitaev2006, Bonderson07} where $\ol{a}$ is the anti-particle of $a$, namely the unique topological charge which satisfies $a\times \ol{a}=I+\cdots$. In our case, the $G$-graded fusion rule implies that $\ol{\sigma_\mb{g}}=\sigma_{\mb{g}^{-1}}$.}
, we must have $\sigma_{\mb{gh}}^+\times {\sigma_{\mb{g}^{-1}}}=2\sigma_\mb{h}$.
The left-hand side has dimension $\sqrt{2}$ while the right-hand already has dimension $2\sqrt{2}$, which is clearly impossible. So we conclude that $\sigma_\mb{g}\times\sigma_\mb{h}=\sigma_{\mb{gh}}^+ + \sigma_{\mb{gh}}^-$.
On the other hand, if the $\mb{g}$ sector has a non-Abelian defect but the $\mb{h}$ sector has Abelian ones, the only available fusion rule is $\sigma_\mb{g}\times\sigma_\mb{h}^\pm=\sigma_{\mb{gh}}$, implying that the defect in the $\mb{gh}$ sector is also non-Abelian. 

What we have just established is that whether the $\mb{g}$-sector is non-Abelian or not gives a homomorphism from $G$ to $\mathbb{Z}_2$. In addition, the fusion rule of the non-Abelian defects implies that they have quantum dimensions $\sqrt{2}$. The inverse statement is quite obvious. Given any such homomorphism, we can write down fusion rules accordingly.
Physically, a non-Abelian defect with quantum dimension $d=\sqrt{2}$ is associated with an odd number of Majorana zero modes localized at the defect~\cite{Kitaev2006}, which implies a topological degeneracy $2^{n-1}$ when there are $2n$ such defects.

We also need to determine the fusion rules of the $G$ sectors consisting of all $\mathbb{Z}_2$-even group elements (i.e. those elements whose defects are Abelian), denoted by $G_\text{e}$ in the following. For any $\mb{g},\mb{h}\in {G}_\text{e}$, we need to specify whether $\sigma_\mb{g}^+\times\sigma_\mb{h}^+$ is $\sigma_{\mb{gh}}^+$ or $\sigma_{\mb{gh}}^-=\psi\times\sigma_{\mb{gh}}^+$. We can generally write
\begin{equation}
	\sigma_\mb{g}^+\times\sigma_\mb{h}^+=\psi^{n(\mb{g,h})}\times\sigma_{\mb{gh}}^+,
	\label{}
\end{equation}
where $n(\mb{g,h})=0, 1$. Since fusion must be associative, comparing $(\sigma_\mb{g}^+\times\sigma_\mb{h}^+)\times\sigma_\mb{k}^+$ and $\sigma_\mb{g}^+\times(\sigma_\mb{h}^+\times\sigma_\mb{k}^+)$ yields
\begin{equation}
	n(\mb{g,h})+n(\mb{gh,k})=n(\mb{g,hk})+n(\mb{h,k})\text{ mod }2.
	\label{}
\end{equation}
Formally, this means that $(-1)^{n(\mb{g,h})}$ is a $\mathbb{Z}_2$-valued $2$-cocycle: $(-1)^{n(\mb{g,h})} \in Z^2(G_\text{e},\mathbb{Z}_2)$. However, we also realize that the definition of $\sigma_\mb{g}^+$ is completely arbitrary and has no physical meaning.  One can always redefine $\tilde{\sigma}_\mb{g}^+=\psi^{m_\mb{g}}\times\sigma_\mb{g}^+$. In terms of $\tilde{\sigma}_\mb{g}^\pm$, we find
\begin{equation}
	\tilde{n}(\mb{g,h})=m_\mb{g}+m_\mb{h}-m_{\mb{g,h}}+n(\mb{g,h})\text{ mod 2}.
	\label{}
\end{equation}
$n(\mb{g,h})$ related by such redefinitions should be considered physically indistinguishable. The equivalence classes are classified by the second group cohomology $H^2(G_\text{e}, \mathbb{Z}_2)$. As we will show later, the cohomology class can also be understood as the projective local symmetry transformations on the fermion-parity  fluxes.

\newsection{Classifying fSPT phases}
As we mentioned before, to get a complete classification of fSPT phases we need to have the algebraic data of the defects, which can be obtained by solving a set of consistency conditions~\cite{BBCW}. Given a particular fusion rule of defects, there may be more than one distinct set of algebraic data. On the other hand, for certain fusion rules it is possible that the consistency conditions do not allow any solutions, in which case the fusion rules do not correspond to any two-dimensional fSPT phases, i.e. there are obstructions~\cite{Chen_arxiv2014, BBCW}.

\subsection{Abelian fSPT Phases}
For simplicity, let us start our analysis from the cases where all the defects are Abelian, i.e. we choose the trivial homomorphism from $G$ to $\mathbb{Z}_2$. Such fSPTs will be referred to as Abelian fSPTs. We have shown that the defect fusion rules in this case correspond to $2$-cocycles $\omega\in H^2(G, \mathbb{Z}_2)$. One can actually systematically solve all the algebraic equations, and it turns out that the sufficient and necessary condition for the existence of a solution can be summarized as follows: Define a $\mathrm{U}(1)$-valued $4$-cocycle
\begin{equation}
	O(\mb{g},\mb{h},\mb{k},\mb{l})=(-1)^{n(\mb{g},\mb{h})n(\mb{k},\mb{l})}.
	\label{eqn:obstruction}
\end{equation}
The obstruction vanishes if and only if $O$ is cohomologically trivial.
We notice that Eq. \eqref{eqn:obstruction} agrees exactly with the result of group super-cohomology~\cite{GuSuperCoho}\footnote{A similar result has been obtained in \Ref{ElsePRB2014} by considering the symmetry transformation on the boundary}.
Following \Ref{GuSuperCoho} we denote the obstruction-free subgroup of $H^2(G, \mathbb{Z}_2)$ by $B{H}^2(G,\mathbb{Z}_2)$, and the group of all Abelian fSPT states by $\mathscr{G}_+$.

We briefly sketch the derivation of \eqref{eqn:obstruction}(See Appendix \ref{app:obstruction} for details). The central quantity responsible for the obstruction is the $F$ symbols of defects, defined diagrammatically as
\begin{equation}
\begin{tikzpicture}[baseline={($ (current bounding box) - (0,10pt) $)},scale=0.25]
	\draw[very thick] (0, 8) node[above]{$\sigma^{\lambda_1}_\mb{g}$} -- (2, 6) ;
	\draw[very thick ]  (2, 6) -- (4, 4);
	\draw[very thick]  (4, 8) node[above]{$\sigma_\mb{h}^{\lambda_2}$}-- (2, 6);
    \draw[very thick] (4,4) -- (6,2) ;
	\draw[very thick](8, 8) node[above]{$\sigma_\mb{k}^{\lambda_3}$}-- (4,4);
\end{tikzpicture}
=F^{\sigma_\mb{g}^{\lambda_1} \sigma_\mb{h}^{\lambda_2}\sigma_\mb{k}^{\lambda_3}}
\begin{tikzpicture}[baseline={($ (current bounding box) - (0,10pt) $)},scale=0.25]
\def\shiftx{15};
\def\shifty{0}
\draw[very thick] (\shiftx+0, \shifty+8) node[above]{$\sigma_\mb{g}^{\lambda_1}$} -- (\shiftx+2, \shifty+6) ;
    \draw[very thick]  (\shiftx+2, \shifty+6) -- (\shiftx+4, \shifty+4);
	\draw[very thick]  (\shiftx+4, \shifty+8) node[above]{$\sigma_\mb{h}^{\lambda_2}$}-- (\shiftx+6, \shifty+6);
    \draw[very thick] (\shiftx+4,\shifty+4) -- (\shiftx+6,\shifty+2) ;
	\draw[very thick](\shiftx+8, \shifty+8) node[above]{$\sigma_\mb{k}^{\lambda_3}$}-- (\shiftx+6, \shifty+6);
	\draw[very thick] (\shiftx+6, \shifty+6) -- (\shiftx+4,\shifty+4);
\end{tikzpicture}.
\end{equation}
They can be thought as the basis transformation for the state space defined by the fusion of three defects $\sigma_\mb{g}^{\lambda_1}\times\sigma_\mb{h}^{\lambda_2}\times \sigma_\mb{k}^{\lambda_3}$.
To a large extent, $F$-symbols can be determined by the consistency conditions that they have to satisfy (known as the Pentagon equations). We obtain the following general parametrization of defect $F$ symbols:
\begin{equation}
	F^{\sigma_\mb{g}^{\lambda_1} \sigma_\mb{h}^{\lambda_2}\sigma_\mb{k}^{\lambda_3}}=\nu(\mb{g},\mb{h},\mb{k})\lambda_1^{n(\mb{h},\mb{k})}.
	\label{eqn:fcocycle}
\end{equation}
Here $\lambda=\pm$ labels the two defects in the same sector, and $\nu$ is a $\mathrm{U}(1)$ 3-cochain to be determined. Plugging \eqref{eqn:fcocycle} into the general Pentagon equation we get 
\begin{equation}
	\di\nu=O,
	\label{}
\end{equation}
which implies that $O$ belongs to the trivial cohomology class in $H^4(G, \mathrm{U}(1))$, thus the obstruction vanishing condition.
Once the obstruction vanishes, different solutions of $\nu$ are given by $3$-cocycles in $H^3(G, \mathrm{U}(1))$. Physically, these solutions correspond to stacking bosonic $G$ SPT phases on top of the fSPT phase~\cite{Chen_arxiv2014, BBCW}.

We can further study the group structure of the Abelian fSPT phases. First of all, we observe from the derivation that $H^3(G, \mathrm{U}(1))$ is a normal subgroup of $\mathscr{G}_+$ with $BH^2(G, \mathbb{Z}_2)$ being the quotient group. Let us denote the group $\mathscr{G}_+$ by $(n, \omega)$ where $n$ is an obstruction-free $2$-cocycle and $[\omega]\in H^3(G, \mathrm{U}(1))$. Consider two such fSPT phases $(n, \omega)$ and $(n', \omega')$. When we stack them on top of each other, the new fSPT phase can be seen to correspond to the $2$-cocycle $n+n'$. One of course expects that $n+n'$ is also obstruction-free, and let us check it explicitly:
\begin{equation}
	\begin{split}
	(-1&)^{(n+n')(\mb{g,h})(n+n')(\mb{k,l})}\\
&=\di(\nu\nu')(-1)^{n(\mb{g,h})n'(\mb{k,l})+ n'(\mb{g,h})n(\mb{k,l})}.
	\end{split}
	\label{}
\end{equation}
We apply the following identity~\cite{Steenrod}:
\begin{equation}
	n(\mb{g,h})n'(\mb{k,l})+ n'(\mb{g,h})n(\mb{k,l})= [\di (n\cup_1 n')](\mb{g,h,k,l}).
	\label{}
\end{equation}
Here a linearized $3$-cochain $n\cup_1 n'$ is defined as
\begin{equation}
	(n\cup_1 n')(\mb{g,h,k})=n(\mb{gh,k})n'(\mb{g,h})+n(\mb{g,hk})n'(\mb{h,k}).
	\label{}
\end{equation}

Therefore we can define the group structure as follows: each phase is mathematically labeled by a triplet $n, \nu_n, \omega$, such that $(\di\nu)(\mb{g,h,k,l})=n(\mb{g,h})n(\mb{k,l})$, and $\omega$ a $3$-cocycle. Notice that of course there are many choices of $\nu$, but one arbitrarily picks one for a fixed $n$ as a reference point, thus our notation $\nu_n$. It is obviously convenient to set $\nu_0=1$. The addition rule we just derived implies that $\nu_{n+n'}$ differs from $\nu_n\nu_{n'}(-1)^{n\cup_1 n'}$ by a $3$-cocycle. Therefore, we should define
\begin{equation}
	\begin{split}
	(n,\nu_n, \omega)&\oplus  (n',\nu_{n'},\omega')=\\
	&\big(n+n', \nu_{n+n'}, \omega \omega' \frac{\nu_n\nu_{n'}}{\nu_{n+n'}}(-1)^{n\cup_1n'}\big).
	\end{split}
	\label{eqn:addition}
\end{equation}
This result was also obtained in \Ref{Bhardwaj2017}.

\subsection{Non-Abelian fSPT Phases}
Next we consider the non-Abelian fSPT phases. First we show that given a nontrivial $\mathbb{Z}_2$ homomorphism of $G$ there exists at least one non-Abelian fSPT, by explicitly constructing the defect theory.  The fusion rules of the $\mathbb{Z}_2$-odd $G$-sectors are fixed by the homomorphism, and we choose a trivial $2$-cocycle in $H^2(G_\mathrm{e}, \mathbb{Z}_2)$, i.e. $n({\mb{g,h}})\equiv 0$ for the fusion rules of the $\mathbb{Z}_2$-even Abelian sectors. We define a map $\varphi$ from the topological charges of the defect theory to those of the familiar Ising topological phase, which has three topological charges $\{\tilde{I}, \tilde{\psi},\tilde{\sigma}\}$:
\begin{equation}
	\varphi(\sigma_\mb{g}^+)= \tilde{I}, \varphi(\sigma_\mb{g}^-)=\tilde{\psi},
	\varphi(\sigma_{\mb{h}})=\tilde{\sigma}.
	\label{}
\end{equation}
All the algebraic data follow from this map and the data of the Ising category~\cite{Kitaev2006}. We will refer to the corresponding fSPT state as the root non-Abelian fSPT state. In fact, such a root fSPT state can be realized with non-interacting fermions: consider a spin-$1/2$ superconductor, where spin $\uparrow$ ($\downarrow$) fermions form a $p_x+ip_y$ ($p_x-ip_y$) superconductor. With a group homomorphism $\rho: G\rightarrow \mathbb{Z}_2=\{0,1\}$, we define the symmetry transformation $R_\mathbf{g}$ on the system as 
\begin{equation}
	R_\mb{g}=
	\begin{cases}
		(-1)^{N_\uparrow} & \rho(\mb{g})=1\\
		\mathds{1} & \rho(\mb{g})=0
	\end{cases}.
	\label{}
\end{equation}
Here $N_\uparrow$ is the number of spin $\uparrow$ fermions.


We now argue other non-Abelian fSPT phases with the same $\mathbb{Z}_2$ homomorphism can all be generated from the root phase.  We make use of the fact that a non-Abelian defect must localize an odd number of Majorana zero modes. Consider two non-Abelian fSPT states ${\mathrm{fSPT}_1}$ and ${\mathrm{fSPT}_2}$ corresponding to the same $\mathbb{Z}_2$ homomorphism of $G$. Denote the sum $\mathrm{fSPT}_3=\mathrm{fSPT}_1+ \mathrm{fSPT}_2$ as the SPT phase obtained by stacking them.  Suppose we create a $\mb{g}$-defect.  If $\mb{g}$ is a $\mathbb{Z}_2$-even element, the defect is already Abelian both in  $\mathrm{fSPT}_1$ and $\mathrm{fSPT}_2$, so is in $\mathrm{fSPT}_3$. if $\mb{g}$ is a $\mathbb{Z}_2$-odd element, because $\mathrm{fSPT}_1$ and $\mathrm{fSPT}_2$ have the same $\mathbb{Z}_2$ homomorphisms the defect localizes an even number of Majorana zero modes altogether, and can only be an Abelian one. Therefore all defects in $\mathrm{fSPT}_3$ are Abelian. It immediately follows that any non-Abelian fSPT state is equivalent to the sum of the ``root'' state of the same $\mathbb{Z}_2$ homomorphism and an Abelian fSPT state. This provides a complete classification of the non-Abelian fSPT phases. 

\subsection{Group structure of fermionic SPT phases}
\label{sec:group-structure}
Now we consider further the Abelian group structure of fSPT phases. First let us set up the notations: we define $\mathscr{G}$ as the Abelian group of all fSPT phases with symmetry $G$, and $\mathscr{G}_+$ as the subgroup of $\mathscr{G}$ consisting of all Abelian fSPT phases. We also denote $\mathscr{G}_B=H^3(G, \mathrm{U}(1))$ as the group of $G$-symmetric bosonic SPT phases.

From our discussion,  we immediately see that $\mathscr{G}_+$ is a normal subgroup of $\mathscr{G}$ with $H^1(G,\mathbb{Z}_2)$ being the quotient group:
\begin{equation}
	\mathscr{G}/\mathscr{G}_+ = H^1(G, \mathbb{Z}_2).
	\label{}
\end{equation}

Furthermore, we see that the root non-Abelian fSPT is essentially a $\mathbb{Z}_2$ fSPT. Notice that given a homomorphism from $G$ to $\mathbb{Z}_2$, we can define a $\mathbb{Z}_2$ $2$-cocycle on $G$ by pulling back the nontrivial $2$-cocycle in $H^2(\mathbb{Z}_2, \mathbb{Z}_2)$. Namely, given a homomorphism $f:G\rightarrow \mathbb{Z}_2=\{0,1\}$, let $\omega(\mb{g,h})=(-1)^{f(\mb{g})f(\mb{h})}$.
We thus conjecture that the addition of two root non-Abelian fSPT with the same homomorphism yields an Abelian fSPT given by this $2$-cocycle $\omega$.

We now consider the group $\mathscr{G}_+$. Again, $\mathscr{G}_B$ is a normal subgroup of $\mathscr{G}_+$:
\begin{equation}
	\mathscr{G}_+/\mathscr{G}_B=BH^2(G, \mathbb{Z}_2).
	\label{}
\end{equation}

The group structure of the Abelian fSPTs has been given in Eq. \eqref{eqn:addition}. For our examples below, it is useful to notice that simplification occurs when summing two identical fSPTs. In this case, $n\cup_1 n=0$, and Eq. \eqref{eqn:addition} simplifies to
\begin{equation}
	(n, \nu, \omega)\oplus (n,\nu,\omega)=(0, 1, \nu^2).
	\label{}
\end{equation}
We can also get the result by directly examing the structure of defect $F$ symbols:  the square of the defect $F$ symbols is $\nu^2$, and $\di( \nu^2)=0$, i.e. $\nu^2\in Z^3(G, \mathrm{U}(1))$. This implies that ``adding up'' two Abelian fSPTs given by the same class in $H^2(G, \mathbb{Z}_2)$ results in a bosonic SPT phase labeled by $\nu^2$~\cite{BifSPT, YoufSPT}.

\newsection{Gauging the fermion parity}
We have established the classification of fSPT phases with unitary symmetries. However, it is clear that the approach can not be extended to anti-unitary symmetries, and we would like to have more direct physical characterization of the SPT phases. In the following we propose an alternative approach to charaterize and classify fSPT phases: we gauge the $\mathbb{Z}_2$ fermion parity symmetry, and we will show that the nontrivial symmetry action on the $\mathbb{Z}_2$ fermion-parity fluxes can be used to characterize fSPT phases.


For fSPT states, it is easy to see that the gauged theory has four anyons $\{I, e, m, \psi\}$, where $m$ is the $\mathbb{Z}_2$ gauge flux, $\psi$ is the fermion, $e=m\times \psi$ can be considered as the (bosonic) $\mathbb{Z}_2$ charge, which is also a $\mathbb{Z}_2$ gauge flux for the $\psi$ fermions. The topological order is identical to that of a $\mathbb{Z}_2$ toric code lattice model. Importantly, the gauged theory preserves the $G$ symmetry and is therefore a $G$-symmetry enriched $\mathbb{Z}_2$ gauge theory.  Therefore, a symmetric adiabatic path between two fSPT phases maps exactly to a symmetric adiabatic path between the corresponding $\mathbb{Z}_2$ topological phases, and consequently if the two symmetry-enriched $\mathbb{Z}_2$ topological phases are distinct, the original fSPTs must be distinct too~\footnote{A caveat here is that two distinct fSPT phases can correspond to the same symmetry-enriched toric code after gauging the fermion parity.}. In this approach we can consider anti-unitary symmetries, or the fermions carrying projective representations of the symmetry group.

Symmetry enrichment in the toric code model can be analyzed using the general theory developed in Ref. \onlinecite{BBCW} ({See also \cite{Fidkowski_unpub} for related discussions}). First of all, one needs to specify the symmetry action on the topological charge labels of anyons. It is easy to see that besides a trivial action, there is a $\mathbb{Z}_2$ action that permutes the $e$ and $m$ particles. These two possible actions on the label set form a $\mathbb{Z}_2$ topological symmetry group. Therefore, the symmetry action on the charge labels is specified by a group homomorphism $\rho$ from $G$ to $\mathbb{Z}_2$. It is clear that this is the same $\mathbb{Z}_2$ homomorphism that classifies the non-Abelian fusion rules of symmetry defects in fSPT states.

Once the symmetry action $\rho$ on charge labels is specified, we can classify patterns of symmetry fractionalization, i.e. anyons carrying projective representations of the symmetry group. This is captured by the group cohomology $H^2_\rho(G, \mathbb{Z}_2\times\mathbb{Z}_2)$, where the subscript $\rho$ indicates that $G$ has a nontrivial action on the coefficients. However, we have an additional restriction that $\psi$ should not carry any nontrivial projective representations (i.e $\psi$ transforms linearly). As we now explain, this leads to a $H^2(G, \mathbb{Z}_2)$ classification.
In the following we will proceed heuristically. We refer the readers to \Ref{BBCW} for more rigorous discussions.

 Let us consider the symmetry action on a general quasiparticle state on a sphere/disk without loss of generality, and assume that the permutation $\rho$ is trivial for simplicity. The global symmetry operator $R_\mb{g}$ can be decomposed into operators localized on each quasiparticle $U_\mb{g}(a)$. They only form projective representations of $G$, i.e. $U_\mb{g}(a)U_\mb{h}(a)=\eta_a(\mb{g},\mb{h})U_\mb{gh}(a)$, however $\eta_a(\mb{g},\mb{h})$ must be consistent with fusion rules:
\begin{equation}
	\eta_a(\mb{g},\mb{h})\eta_b(\mb{g},\mb{h})=\eta_c(\mb{g},\mb{h}), \text{if }N_{ab}^c>0.
	\label{eqn:char}
\end{equation}
This stems from the fact that $R_\mb{g}R_\mb{h}=R_\mb{gh}$ must hold on vaccume state.
In particular, we have $\eta_e^2=\eta_m^2=1, \eta_\psi=\eta_e\eta_m$.

On the other hand, from the associativity of operator products $U_\mb{g}U_\mb{h}U_\mb{k}$, we have
\begin{equation}
	\eta_a(\mb{h},\mb{k})\eta_a(\mb{g},\mb{hk})=\eta_a(\mb{g},\mb{h})\eta_a(\mb{gh},\mb{k}).
	\label{}
\end{equation}
So naively, one may conclude that gauge-inequivalent classes of $\eta$ are classified by $H^2(G, \mathbb{Z}_2\times\mathbb{Z}_2)$. However, 
since we are considering symmetry-enriched toric code from gauging a fSPT phase, the $\psi$ quasiparticle has to form a linear representation, implying $\eta_\psi$ can be chosen to $1$ and therefore $\eta_e=\eta_m$. So the actually classification is just $H^2(G, \mathbb{Z}_2)$.

We also notice that any Abelian phase on anyon $a$ that satisfies a relation like \eqref{eqn:char} must be the braiding phase of a (fixed) Abelian anyon with $a$, which actually holds for any topological phase following from modularity. The fact that we have $\eta_e=\eta_m$ means $\eta_e$ ($\eta_m$) is the braiding phase of either $I$ or $\psi$ with $e$($m$). Physically, we can think of $U_\mb{g}(a)$ as taking the $\sigma_{\mb{g}}^+$ defect around $a$. If $\eta_e=\eta_m=-1$, the only possiblity is that when $\sigma_{\mb{g}}^+$ fuses with $\sigma_\mb{h}^+$, we obtain $\sigma_{\mb{gh}}^+$ and also a fermion $\psi$ which when taking a full braid around $e$($m$) generates the phases $\eta_e$($\eta_m$). Therefore, if we write $\eta_e(\mb{g},\mb{h})=\eta_m(\mb{g},\mb{h})=(-1)^{n(\mb{g},\mb{h})}$ with $n(\mb{g},\mb{h})=0,1$ being the linearized $\mathbb{Z}_2$ $2$-cocycle, we have
\begin{equation}
	\sigma_\mb{g}^+\times\sigma_\mb{h}^+=\psi^{n(\mb{g},\mb{h})}\sigma_{\mb{gh}}^+.
	\label{}
\end{equation}

Therefore we recover the previous classification. We notice that this also provides a physical characterization of $H^2(G, \mathbb{Z}_2)$, through local projective symmetry actions on the $\mathbb{Z}_2$ fermion parity flux in a fermionic system, which can be measured in numerical simulations~\cite{PollmannSET, ZaletelPRB2014}.

\newsection{Examples}
In this section we apply the general theory to $G=\mathbb{Z}_n, \mathbb{Z}_2\times\mathbb{Z}_2$ and $\mathbb{Z}_2^T$.

\subsection{$G=\mathbb{Z}_n$}
We label the group elements of $\mathbb{Z}_n$ by $a=0,1,\dots, n-1$ and the group multiplication is written additively, i.e. $a+b=[a+b]$ where $[a]$ is $a$ mod $n$.
First we have $H^1(\mathbb{Z}_n, \mathbb{Z}_2)=H^2(\mathbb{Z}_n,\mathbb{Z}_2)=\mathbb{Z}_{(n,2)}$. For odd $n$, there are only bosonic SPT phases classified by $H^3(\mathbb{Z}_n, \mathrm{U}(1))=\mathbb{Z}_n$.

For even $n$, first of all there is a unique homomorphism from $\mathbb{Z}_n$ to $\mathbb{Z}_2$, namely $f:a\rightarrow (-1)^a$. Applying the rule developed in Sec. \ref{sec:group-structure}, the ``square'' of the root non-Abelian fSPT has $\omega(a,b)=(-1)^{ab}$.  

For the Abelian fSPTs, the nontrivial $2$-cocycle in $H^2(\mathbb{Z}_n,\mathbb{Z}_2)$ is given by 
\begin{equation}
\omega(a,b) = e^{\frac{i\pi}{n}([a]+[b]-[a+b])}.
	\label{}
\end{equation}
	To see this is a nontrivial cocycle, it suffices to notice that $\omega([\frac{n}{2}],[\frac{n}{2}])=-1$ is a gauge-invariant quantity since we are considering $\mathbb{Z}_2$ coefficients.
Since $H^4(\mathbb{Z}_n, \mathrm{U}(1))$ is trivial, there are no obstructions.  
We can also explicitly find the fermionic $3$-cocycle:
	$\nu(a,b,c)=e^{\frac{i\pi}{n^2}a([b]+[c]-[b+c])}$.
We notice that $\nu(a,b,c)$ is the square root of the generating $\mathrm{U}(1)$ $3$-cocycle in $H^3(\mathbb{Z}_N, \mathrm{U}(1))=\mathbb{Z}_N$. This implies that two fSPT corresponding to the nontrivial $2$-cocycle can be stacked to form the generating bosonic SPT. Therefore, the Abelian fSPT phases with $\mathbb{Z}_n$ symmetry form a $\mathbb{Z}_{2n}$ group. 

Back to the non-Abelian fSPT. By comparing the $2$-cocycles we see that for $n\equiv 2\,(\text{mod }4)$, the square of the non-Abelian fSPT yields the nontrivial Abelian fSPT. Otherwise if $n\equiv 0\,(\text{mod }4)$, the result is a bosonic SPT. We can further fix the bosonic SPT by using the method of anyon condensation, the details of which will be reported elsewhere.

Using these results, we completely determine the group structure of fSPT phases with $G=\mathbb{Z}_n$:
\begin{equation}
	\mathscr{G}=
	\begin{cases}
		\mathbb{Z}_n & n\text{ is odd}\\
		\mathbb{Z}_2\times\mathbb{Z}_{2n} &  n\equiv 0\,(\text{mod }4)\\
		\mathbb{Z}_{4n} & n\equiv 2\,(\text{mod }4)
	\end{cases}.
	\label{}
\end{equation}

Physically, the nontrivial Abelian fSPT is characterized by the ``half'' $\mathbb{Z}_N$ charge of the fermion parity flux. Alternatively, we have the following fusion rules for $\mathbb{Z}_N$ defects:
	\begin{equation}
		\sigma_{[1]}^N = \psi.
		\label{eqn:ZNdefectfusion}
	\end{equation}
	Here $[1]$ denotes the generator of the $\mathbb{Z}_N$ group.

	Let us also discuss physical realizations of these fermionic SPT phases. A model for the ``root'' non-Abelian SPT phase has been presented in. For the nontrivial Abelian SPT phase, we can consider a model of spin-$1/2$ fermions, where spin-up (down) fermions form a Chern insulator with Chern number $1$ ($-1$). In the presence of $S_z$ conservation, one can say that the spin Chern number is $1$. The model actually has $\mathrm{U}(1)_\uparrow \times \mathrm{U}(1)_\downarrow$ symmetries. Now we break $\mathrm{U}(1)_\uparrow$ down to $\mathbb{Z}_N$, i.e. number of spin-up fermions only conserved up to $N$.
	The defect fusion rule  Eq. \eqref{eqn:ZNdefectfusion} can be understood using the spin Hall response: a $\mathbb{Z}_N$ symmetry defect is nothing but a $\frac{2\pi}{N}$ flux of $\mathrm{U}(1)_\uparrow$. Spin up fermions have $\mathrm{U}(1)_\uparrow$ Hall conductance $\sigma_H=\frac{e^2}{h}$, which means that a $\frac{2\pi}{N}$ flux must trap $\frac{1}{N}$ electric charge. Therefore, if we insert $N\cdot \frac{2\pi}{N}=2\pi$ flux, a unit charge is accumulated corresponding to a fermion, which is Eq. \eqref{eqn:ZNdefectfusion}. This is of course nothing but Laughlin's famous argument.

\subsection{$G=\mathbb{Z}_2\times\mathbb{Z}_2$ }
We can easily see that $H^1(G, \mathbb{Z}_2)=\mathbb{Z}_2^2$, corresponding to non-Abelian fSPTs protected by any of the three  $\mathbb{Z}_2$ subgroups.

We now determine the group structure for $G=\mathbb{Z}_2\times\mathbb{Z}_2=\{1, X, Y, XY\}$. We have $H^1(G, \mathbb{Z}_2)=\mathbb{Z}_2^2$. The three nontrivial classes can be understood as $X$ being odd, or $Y$ being odd. Then $H^2(G, \mathbb{Z}_2)=\mathbb{Z}_2^3$. The three generating classes can be labeled by 1) $\omega(X,X)=\omega(XY, XY)=-1, \omega(Y,Y)=1$, (2) $\omega(Y,Y)=\omega(XY, XY)=-1$, $\omega(X,X)=1$ and (3) $\omega(X,Y)/\omega(Y,X)=-1$. The first two only require $X$ (or $Y$) to be nontrivial, and as we already explained for $\mathbb{Z}_2$ symmetry the fermion-parity flux carries a half $\mathbb{Z}_2$ charge for the subgroup generated by $X$ (or $Y$). Therefore combined with $H^1$ classes and the bosonic SPT phases, we get a $\mathbb{Z}_8\times \mathbb{Z}_8$ classification. The more interesting Abelian fSTP, labeled by $\omega(X,Y)/\omega(Y,X)=-1$, is physically characterized by an irreducible $2$-dimensional representation on the fermion-parity $\pi$ flux. This part gives $\mathbb{Z}_4$ classification. In summary, we find $\mathbb{Z}_8\times\mathbb{Z}_8\times\mathbb{Z}_4$ classification for $G=\mathbb{Z}_2\times\mathbb{Z}_2$.


\subsection{$G=\mathbb{Z}_2^T=\{1,T\}$}
\label{sec:Z2T}
We now discuss time-reversal symmetry. Our method of symmetry extension does not apply because the time-reversal symmetry is anti-unitary. However, we can still consider gauging the fermion parity and study the symmetry action in the gauged model. Since after gauging one has a bosonic Hilbert space, the time-reversal symmetry operator satisfies $T^2=1$. We can then distinguish two cases, where the fermion $\psi$ is a Kramers singlet ($T^2=1$) or a doublet ($T^2=-1$).

First let us specify the symmetry action $\rho$ on the charge labels. There are two possibilities: (a) $T$ does not change charge labels at all. (b) $T$ exchanges $e$ and $m$. Interestingly, in the latter case $\psi$ must have $T^2=-1$~\cite{metlitski2013, Fidkowski_PRX2013}. Therefore we immediately see that there is a 2D fSPT with $T^2=-1$ fermions, in which the local fermion parity of a $\pi$ vortex changes under the time-reversal operation~\cite{QiPRL2009, GuNeutrino, Metlitski_arxiv2014}, and there are no other symmetry fractionalization classes due to $H^2_\rho(\mathbb{Z}_2^T, \mathbb{Z}_2\times\mathbb{Z}_2)=\mathbb{Z}_1$.

Let us consider the symmetry fractionalization class of the trivial action on the charge labels, which is classified by $H^2(\mathbb{Z}_2^T, \mathbb{Z}_2\times\mathbb{Z}_2)=\mathbb{Z}_2\times\mathbb{Z}_2$. Physically, the four classes correspond to four possible ways of assigning $T^2=\pm 1$ to the four charges. For the two classes with $\psi$ being a Kramers doublet, one of the $e$ or $m$ charges has to be a Kramers singlet, which means that the $\pi$ vortex is trivial in the fSPT. Therefore they do not correspond to any nontrivial fSPT phases. We are then left with one nontrivial fractionalization class with $\psi$ being a Kramers singlet, and both $e$ and $m$ being Kramers doublets. However, this fractionalization class is known to be anomalous, i.e. there is an obstruction to realize it in two dimensions~\cite{senthil_3D, WangPRB2013} and thus does not correspond to a fSPT in 2D. Together with the fact that $H^3(\mathbb{Z}_2^T, \mathrm{U}(1))=\mathbb{Z}_1$, we conclude that there is only one nontrivial 2D fSPT phase with $\mathbb{Z}_2^T$ symmetry, the class DIII topological superconductor.

\newsection{Acknowledgement} We thank Zhenghan Wang and Daniel Freed for enlightening discussions. M.C. would like to thank Parsa Bonderson and Maissam Barkeshli for collaboration on related projects, and Perimeter Institute for hospitality where part of the work was done. Research at Perimeter Institute is supported by the Government of Canada through Industry Canada and by the Province of Ontario through the Ministry of Research.  Z.C.G also acknowledges start up support from Department of Physics, The Chinese University of Hong Kong, Direct Grant No. 4053224 from The Chinese University of Hong Kong and the funding from RGC/ECS(No.2191110).

\appendix
\section{Review of Group Cohomology}
\label{app:cohomology}

In this section, we provide a brief review of group cohomology for finite groups.
Given a finite group $G$, let $M$ be an Abelian group equipped with a $G$ action $\rho: G \rightarrow \text{Aut}(M)$, which is compatible with group multiplication. In particular, for any $\mathbf{g}\in G$ and $a,b \in M$, we have
\begin{equation}
\rho_\mathbf{g}(ab)=\rho_\mathbf{g}(a) \rho_\mathbf{g}(b).
\label{}
\end{equation}
(We leave the group multiplication symbols implicit.) Such an Abelian group $M$ with $G$ action $\rho$ is called a $G$-module.

Let $\omega(\mathbf{g}_1, \dots,\mathbf{g}_n)\in M$ be a function of $n$ group elements $\mathbf{g}_j \in G$ for $j=1,\dots,n$. Such a function is called a $n$-cochain and the set of all $n$-cochains is denoted as $C^n(G, M)$. They naturally form a group under multiplication,
\begin{equation}
  (\omega\cdot\omega')(\mb{g}_1, \dots, \mb{g}_n)=\omega(\mb{g}_1, \dots, \mb{g}_n)\omega'(\mb{g}_1, \dots, \mb{g}_n),
  \label{}
\end{equation}
and the identity element is the trivial cochain $\omega(\mb{g}_1,\dots,\mb{g}_n)=1$.

We now define the ``coboundary'' map $\mathrm{d}: C^n(G, M) \rightarrow C^{n+1}(G, M)$ acting on cochains to be
\begin{equation}
\begin{split}
\mathrm{d}\omega (&\mathbf{g}_1,\dots,\mathbf{g}_{n+1})= \rho_{\mathbf{g}_1}[\omega(\mathbf{g}_2,\dots,\mathbf{g}_{n+1})] 
	\\
	&\times \prod_{j=1}^n \omega^{(-1)^j}(\mathbf{g}_1,\dots,\mathbf{g}_{j-1},\mathbf{g}_j\mathbf{g}_{j+1},\mathbf{g}_{j+1},\dots,\mathbf{g}_{n+1}) \\
    &\times \omega^{(-1)^{n+1}}(\mathbf{g}_1,\dots,\mathbf{g}_{n})
.
\end{split}
\label{}
\end{equation}
One can directly verify that $\mathrm{d} \mathrm{d}\omega=1$ for any $\omega \in C^n(G, M)$, where $1$ is the trivial cochain in $C^{n+2}(G, M)$. This is why $\mathrm{d}$ is considered a ``boundary operator.''

With the coboundary map, we next define $\omega\in C^n(G, M)$ to be an $n$-cocycle if it satisfies the condition $\mathrm{d}\omega=1$. We denote the set of all $n$-cocycles by
\begin{equation}
\begin{split}
Z^n_{\rho}(G, M) &= \text{ker}[\mathrm{d}: C^n(G, M) \rightarrow C^{n+1}(G, M)]  \\
 &= \{ \, \omega\in C^n(G, M) \,\, | \,\, \mathrm{d}\omega=1 \, \}.
\end{split}
\label{}
\end{equation}
We also define $\omega\in C^n(G, M)$ to be an $n$-coboundary if it satisfies the condition $\omega= \mathrm{d} \mu $ for some $(n-1)$-cochain $\mu \in C^{n-1}(G, M)$. We denote the set of all $n$-coboundaries by
Also we have
\begin{equation}
\begin{split}
 B^n_{\rho}(G, M)  &= \text{im}[ \mathrm{d}: C^{n-1}(G, M) \rightarrow C^{n}(G, M) ] \\
 &=\{ \, \omega\in C^n(G, M) \,\, | \,\, \exists \mu \in C^{n-1}(G, M) : \omega = \mathrm{d}\mu \, \}
.
\end{split}
\label{}
\end{equation}

Clearly, $B^n_{\rho}(G, M) \subset Z^n_{\rho}(G, M) \subset C^n(G, M)$. In fact, $C^n$, $Z^n$, and $B^n$ are all groups and the co-boundary maps are homomorphisms. It is easy to see that $B^n_{\rho}(G, M)$ is a normal subgroup of $Z^n_{\rho}(G, M)$. Since d is a boundary map, we think of the $n$-coboundaries as being trivial $n$-cocycles, and it is natural to consider the quotient group
\begin{equation}
H^n_{\rho}(G, M)=\frac{Z^n_{\rho}(G, M)}{B^n_{\rho}(G, M)}
,
\label{}
\end{equation}
which is called the $n$-th group cohomology. In other words, $H^n_{\rho}(G, M)$ collects the equivalence classes of $n$-cocycles that only differ by $n$-coboundaries.

\section{Obstruction to Abelian $G$-crossed Extensions}
\label{app:obstruction}
In this section we derive the obstruction to a consistent Abelian $G$-extension.
We first briefly review the algebraic theory of symmetry defects, known as the $G$-crossed braided extension of a braided tensor category~\cite{BBCW}. For simplicity, we assume all defects (as well as anyons in the original theory) are abelian. The collection of all defects is called the $G$-extension:
\begin{equation}
	\mathcal{C}_G=\bigoplus_{\mb{g}\in G}\mathcal{C}_\mb{g}.
	\label{}
\end{equation}

\begin{figure}[htpb]
	\centering
	\includegraphics[width=0.4\textwidth]{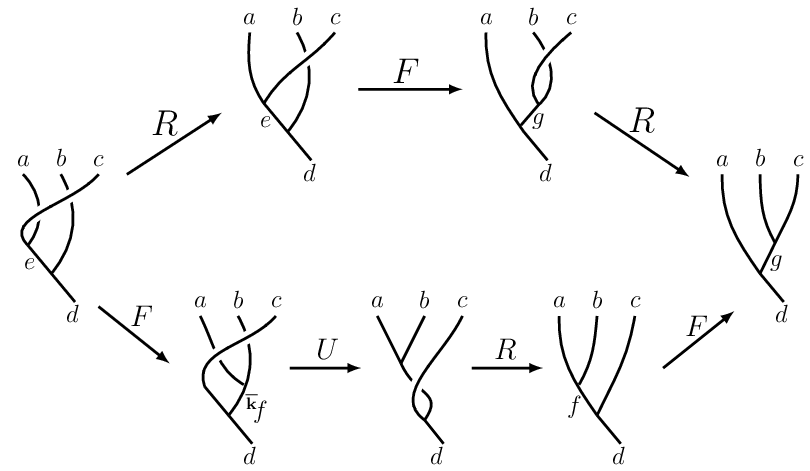}
	\includegraphics[width=0.4\textwidth]{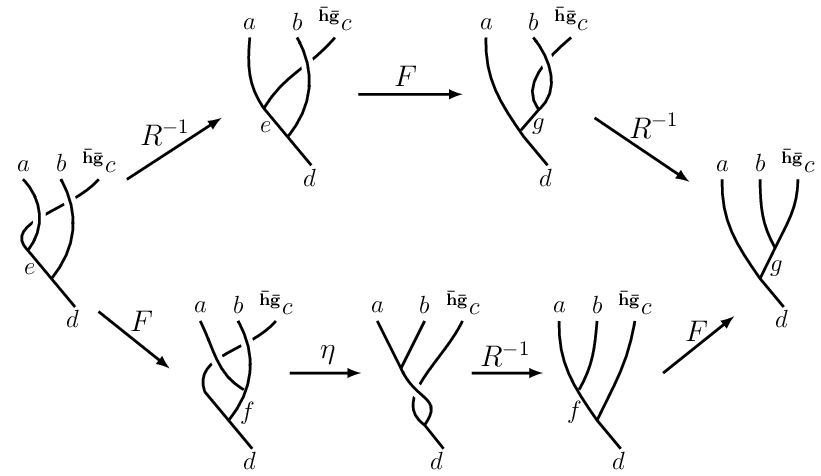}
	\caption{Heptagon equations which ensure consistency between $G$-crossed braiding and fusion of defects.}
	\label{fig:heptagon}
\end{figure}

We will use the diagrammatic formulation (for a review, see \Refs{BBCW, Kitaev2006}). The basic data of the $G$-extension includes:
\begin{itemize}
	\item $G$-graded fusion rules, i.e. $a_\mb{g}\times b_\mb{h}=\sum_{c_{\mb{gh}}}N_{a_\mb{g}b_\mb{h}}^{c_\mb{gh}}c_\mb{gh}$.
	\item $F$ symbols for associativity of fusion.
		\begin{equation}
  \pspicture[shift=-1.0](0,-0.45)(1.8,1.8)
  \small
  \psset{linewidth=0.9pt,linecolor=black,arrowscale=1.5,arrowinset=0.15}
  \psline(0.2,1.5)(1,0.5)
  \psline(1,0.5)(1,0)
  \psline(1.8,1.5) (1,0.5)
  \psline(0.6,1) (1,1.5)
   \psline{->}(0.6,1)(0.3,1.375)
   \psline{->}(0.6,1)(0.9,1.375)
   \psline{->}(1,0.5)(1.7,1.375)
   \psline{->}(1,0.5)(0.7,0.875)
   \psline{->}(1,0)(1,0.375)
   \rput[bl]{0}(0.03,1.6){$a_\mb{g}$}
   \rput[bl]{0}(0.92,1.6){$b_\mb{h}$}
   \rput[bl]{0}(1.72,1.6){${c}_\mb{k}$}
   \rput[bl]{0}(0.5,0.5){$e$}
   \rput[bl]{0}(0.9,-0.35){$d_\mb{ghk}$}
 \scriptsize
   \rput[bl]{0}(0.3,0.8){$\alpha$}
   \rput[bl]{0}(0.7,0.25){$\beta$}
  \endpspicture
  = \sum_{f,\mu,\nu} \left[F_{d_{\mb{ghk}}}^{a_\mb{g}b_\mb{h}c_\mb{k}}\right]_{(e,\alpha,\beta)(f,\mu,\nu)}
 \pspicture[shift=-1.0](0,-0.45)(1.8,1.8)
  \small
  \psset{linewidth=0.9pt,linecolor=black,arrowscale=1.5,arrowinset=0.15}
  \psline(0.2,1.5)(1,0.5)
  \psline(1,0.5)(1,0)
  \psline(1.8,1.5) (1,0.5)
  \psline(1.4,1) (1,1.5)
   \psline{->}(0.6,1)(0.3,1.375)
   \psline{->}(1.4,1)(1.1,1.375)
   \psline{->}(1,0.5)(1.7,1.375)
   \psline{->}(1,0.5)(1.3,0.875)
   \psline{->}(1,0)(1,0.375)
   \rput[bl]{0}(0.05,1.6){$a_\mb{g}$}
   \rput[bl]{0}(0.90,1.6){$b_\mb{h}$}
   \rput[bl]{0}(1.75,1.6){${c}_\mb{k}$}
   \rput[bl]{0}(1.25,0.45){$f$}
   \rput[bl]{0}(0.9,-0.35){$d_{\mb{ghk}}$}
 \scriptsize
   \rput[bl]{0}(1.5,0.8){$\mu$}
   \rput[bl]{0}(0.7,0.25){$\nu$}
  \endpspicture
.
\label{eqn:fsymbol}
\end{equation}
The $F$ symbols can be viewed as changes of bases for the states associated with quasiparticles.

\item $G$ action on labels, defined by maps $\rho_\mb{g}$, which acts on the topological charge labels in the following way: $\rho_\mb{h}(a_\mb{g})\in \mathcal{C}_{\mb{hgh^{-1}}}$. We write $\rho_\mb{h}(a_\mb{g})$ as $^\mb{h}a_\mb{g}$.
Diagrammatically, this means that when topological charge lines cross, the labels should change accordingly:
	\begin{equation}
	\pspicture[shift=-0.75](-0.1,-0.4)(1.3,1.4)
\small
  \psset{linewidth=0.9pt,linecolor=black,arrowscale=1.5,arrowinset=0.15}
  \psline(0.96,0.05)(0.2,1)
  \psline{->}(0.96,0.05)(0.28,0.9)
  \psline(0.24,0.05)(1,1)
  \psline[border=2pt]{->}(0.24,0.05)(0.92,0.9)
  \rput[bl]{0}(-0.1,1.1){$^{\bf h}a_{\bf g}$}
  \rput[br]{0}(1.2,1.1){$b_{\bf h}$}
  \rput[bl]{0}(-0.15,-0.05){$b_{\bf h}$}
  \rput[br]{0}(1.45,-0.1){$ a_{\bf g}$}
  \endpspicture
		\label{}
	\end{equation}
In our derivation we just need the obvious fact that $\rho_1(a_\mb{g})=a_\mb{g}$.

\item $G$ action on fusion spaces, defined by unitary transformations $U_\mb{k}(a_\mb{g}, b_\mb{h}; c_{\mb{gh}})$:
	\begin{equation}
\label{eq:GcrossedU}
\psscalebox{.8}{
\pspicture[shift=-1.7](-0.8,-0.8)(1.8,2.4)
  \small
  \psset{linewidth=0.9pt,linecolor=black,arrowscale=1.5,arrowinset=0.15}
  \psline{->}(0.7,0)(0.7,0.45)
  \psline(0.7,0)(0.7,0.55)
  \psline(0.7,0.55)(0.25,1)
  \psline(0.7,0.55)(1.15,1)	
  \psline(0.25,1)(0.25,2)
  \psline{->}(0.25,1)(0.25,1.9)
  \psline(1.15,1)(1.15,2)
  \psline{->}(1.15,1)(1.15,1.9)
  \psline[border=2pt](-0.65,0)(2.05,2)
  \psline{->}(-0.65,0)(0.025,0.5)
  \rput[bl]{0}(-0.4,0.6){$x_{\bf k}$}
  \rput[br]{0}(1.5,0.65){$^{\bf \bar k}b$}
  \rput[bl]{0}(0.4,-0.4){$^{\bf \bar k}c_{\bf gh}$}
  \rput[br]{0}(1.3,2.1){$b_{\bf h}$}
  \rput[bl]{0}(0.0,2.05){$a_{\bf g}$}
 \scriptsize
  \rput[bl]{0}(0.85,0.35){$\mu$}
  \endpspicture
}
=
\sum_{\nu} \left[ U_{\bf k}\left(a, b; c\right)\right]_{\mu \nu}
\psscalebox{.8}{
\pspicture[shift=-1.7](-0.8,-0.8)(1.8,2.4)
  \small
  \psset{linewidth=0.9pt,linecolor=black,arrowscale=1.5,arrowinset=0.15}
  \psline{->}(0.7,0)(0.7,0.45)
  \psline(0.7,0)(0.7,1.55)
  \psline(0.7,1.55)(0.25,2)
  \psline{->}(0.7,1.55)(0.3,1.95)
  \psline(0.7,1.55) (1.15,2)	
  \psline{->}(0.7,1.55)(1.1,1.95)
  \psline[border=2pt](-0.65,0)(2.05,2)
  \psline{->}(-0.65,0)(0.025,0.5)
  \rput[bl]{0}(-0.4,0.6){$x_{\bf k}$}
  \rput[bl]{0}(0.4,-0.4){$^{\bf \bar k}c_{\bf gh}$}
  \rput[bl]{0}(0.15,1.2){$c_{\bf gh}$}
  \rput[br]{0}(1.3,2.1){$b_{\bf h}$}
  \rput[bl]{0}(0.0,2.05){$a_{\bf g}$}
 \scriptsize
  \rput[bl]{0}(0.85,1.35){$\nu$}
  \endpspicture
}
.
\end{equation}
Importantly, we have the normalization condition $U_1\equiv \openone$.

\item Natural isomorphisms $\eta_{x_\mb{k}}(\mb{g},\mb{h})$ on topological charges, which define the projective $G$ actions:
	\begin{equation}
	\psscalebox{.8}{
\pspicture[shift=-1.7](-0.8,-0.8)(1.8,2.4)
  \small
  \psset{linewidth=0.9pt,linecolor=black,arrowscale=1.5,arrowinset=0.15}
  \psline(-0.65,0)(2.05,2)
  \psline[border=2pt](0.7,0.55)(0.25,1)
  \psline[border=2pt](1.15,1)(1.15,2)
  \psline(0.7,0.55)(1.15,1)	
  \psline{->}(0.7,0)(0.7,0.45)
  \psline(0.7,0)(0.7,0.55)
  \psline(0.25,1)(0.25,2)
  \psline{->}(0.25,1)(0.25,1.9)
  \psline{->}(1.15,1)(1.15,1.9)
  \psline{->}(-0.65,0)(0.025,0.5)
  \rput[bl]{0}(-0.5,0.6){$x_{\bf k}$}
  \rput[bl]{0}(0.5,1.3){$^{\bf \bar g}x$}
  \rput[bl]{0}(1.6,2.1){$^{\bf \bar h \bar g}x_{\bf k}$}
  \rput[bl]{0}(0.4,-0.35){$c_{\bf gh}$}
  \rput[br]{0}(1.3,2.1){$b_{\bf h}$}
  \rput[bl]{0}(0.0,2.05){$a_{\bf g}$}
 \scriptsize
  \rput[bl]{0}(0.85,0.35){$\mu$}
  \endpspicture
}
=
\eta_{x}\left({\bf g},{\bf h}\right)
\psscalebox{.8}{
\pspicture[shift=-1.7](-0.8,-0.8)(1.8,2.4)
  \small
  \psset{linewidth=0.9pt,linecolor=black,arrowscale=1.5,arrowinset=0.15}
  \psline(-0.65,0)(2.05,2)
  \psline[border=2pt](0.7,0)(0.7,1.55)
  \psline{->}(0.7,0)(0.7,0.45)
  \psline(0.7,1.55)(0.25,2)
  \psline{->}(0.7,1.55)(0.3,1.95)
  \psline(0.7,1.55) (1.15,2)	
  \psline{->}(0.7,1.55)(1.1,1.95)
  \psline{->}(-0.65,0)(0.025,0.5)
  \rput[bl]{0}(-0.5,0.6){$x_{\bf k}$}
  \rput[bl]{0}(1.6,2.1){$^{\bf \bar h \bar g}x_{\bf k}$}
  \rput[bl]{0}(0.4,-0.35){$c_{\bf gh}$}
  \rput[br]{0}(1.3,2.1){$b_{\bf h}$}
  \rput[bl]{0}(0.0,2.05){$a_{\bf g}$}
 \scriptsize
  \rput[bl]{0}(0.85,1.35){$\mu$}
  \endpspicture
}	
		\label{}
	\end{equation}

	Similarly, $\eta$ are normalized: $\eta_x(1, \mb{h})=\eta_x(\mb{g},1)=1$.

\item $G$-crossed $R$ symbols, defined by the following diagrammatic relation:
\begin{equation}
\pspicture[shift=-0.8](-0.1,-0.5)(1.5,1.4)
  \small
  \psset{linewidth=0.9pt,linecolor=black,arrowscale=1.5,arrowinset=0.15}
  \psline{->}(0.7,0)(0.7,0.43)
  \psline(0.7,0)(0.7,0.5)
 \psarc(0.8,0.6732051){0.2}{120}{240}
 \psarc(0.6,0.6732051){0.2}{-60}{35}
  \psline (0.6134,0.896410)(0.267,1.09641)
  \psline{->}(0.6134,0.896410)(0.35359,1.04641)
  \psline(0.7,0.846410) (1.1330,1.096410)	
  \psline{->}(0.7,0.846410)(1.04641,1.04641)
  \rput[bl]{0}(0.5,-0.3){$c_{\bf gh}$}
  \rput[br]{0}(1.4,1.15){$b_{\bf h}$}
  \rput[bl]{0}(-0.1,1.15){$a_{\bf g}$}
 \scriptsize
  \rput[bl]{0}(0.82,0.35){$\mu$}
  \endpspicture
=\sum\limits_{\nu }\left[R_{c_{\bf gh}}^{a_{\bf g} b_{\bf h}}\right] _{\mu \nu }
\pspicture[shift=-0.8](-0.1,-0.5)(1.5,1.4)
  \small
  \psset{linewidth=0.9pt,linecolor=black,arrowscale=1.5,arrowinset=0.15}
  \psline{->}(0.7,0)(0.7,0.45)
  \psline(0.7,0)(0.7,0.55)
  \psline(0.7,0.55) (0.25,1)
  \psline{->}(0.7,0.55)(0.3,0.95)
  \psline(0.7,0.55) (1.15,1)	
  \psline{->}(0.7,0.55)(1.1,0.95)
  \rput[bl]{0}(0.5,-0.3){$c_{\bf gh}$}
  \rput[br]{0}(1.4,1.05){$b_{\bf h}$}
  \rput[bl]{0}(-0.1,1.05){$a_{\bf g}$}
 \scriptsize
  \rput[bl]{0}(0.82,0.37){$\nu$}
  \endpspicture
.
\end{equation}%

\end{itemize}
These data satisfy a set of coherence conditions. For our purpose, 
we have to solve the pentagon equations and a generalization of hexagon equations, called heptagon equations (see Fig. \ref{fig:heptagon}), to find consistent $F$ and $R$ symbols.

 The pentagon equation leads to the following general parametrization of $F$ symbols: 
\begin{equation}
	F^{\sigma_\mb{g}^{\lambda_1}\sigma_\mb{h}^{\lambda_2}\sigma_\mb{k}^{\lambda_3}}=\nu(\mb{g},\mb{h},\mb{k})\theta^{\frac{1-\lambda_1}{2}}(\mb{h},\mb{k}).
	\label{}
\end{equation}
Notice that because we are considering Abelian fusion rules, all labels in the definition \eqref{eqn:fsymbol} of $F$ symbol are uniquely determined by the three outgoing lines and therefore suppressed here.
$\nu(\mb{g},\mb{h},\mb{k})$ is a $\mathrm{U}(1)$ $3$-cochain, and $\theta$ is a $\mathbb{Z}_2$ $2$-cocycle. And they should satisfy
\begin{equation}
	(\mathrm{d}\nu)(\mb{g},\mb{h},\mb{k},\mb{l})=\theta^{n_{\mb{g},\mb{h}}}(\mb{k},\mb{l}).
	\label{}
\end{equation}
This implies that the right-hand side must be a $4$-coboundary.

We now use $G$-crossed heptagon equations to find $\theta$. 

Let us first consider the heptagon equation with the three outgoing lines being $\sigma_\mb{g}^{\lambda}, \psi$ and $\psi$ (from left to right).
\begin{equation}
	R^{\sigma_\mb{g}^{\lambda}\psi}F^{\sigma_\mb{g}^{\lambda}\psi\psi}R^{\psi\psi}=F^{\psi\sigma_\mb{g}^{\lambda}\psi}R^{\sigma_\mb{g}^\lambda\times\psi,\psi}F^{\sigma_\mb{g}^{\lambda}\psi\psi}.
	\label{}
\end{equation}
Since $R^{\psi\psi}=-1$, we have
\begin{equation}
	R^{\sigma_\mb{g}^-\psi}=-R^{\sigma_\mb{g}^+\psi}.
	\label{}
\end{equation}

Let us consider the heptagon equation with the three outgoing lines being $\sigma_\mb{g}^{\lambda_1}, \sigma_\mb{h}^{\lambda_2}$ and $\psi$ (from left to right). Because $\psi\in \mathcal{C}_1$, the action on the vertex is trivial. We have
\begin{equation}
	R^{\sigma_\mb{g}^{\lambda_1}\psi}F^{\sigma_\mb{g}^{\lambda_1}\psi\sigma_\mb{h}^{\lambda_2}}R^{\sigma_\mb{h}^{\lambda_2}\psi}=F^{\psi\sigma_\mb{g}^{\lambda_1}\sigma_\mb{h}^{\lambda_2}}R^{\sigma_\mb{g}^{\lambda_1}\times\sigma_\mb{h}^{\lambda_2}, \psi}F^{\sigma_\mb{g}^{\lambda_1}\sigma_\mb{h}^{\lambda_2}\psi}
	\label{}
\end{equation}
Following the general parametrization and the normalization condition of $\nu$, we have $F^{\sigma_\mb{g}^{\lambda_1}\psi\sigma_\mb{h}^{\lambda_2}}=F^{\sigma_\mb{g}^{\lambda_1}\sigma_\mb{h}^{\lambda_2}\psi}=1$ and $F^{\psi\sigma_\mb{g}^{\lambda_1}\sigma_\mb{h}^{\lambda_2}}=\theta(\mb{g},\mb{h})$. Therefore
\begin{equation}
	\theta(\mb{g},\mb{h})=\frac{R^{\sigma_\mb{g}^{\lambda_1}\psi}R^{\sigma_\mb{h}^{\lambda_2}\psi}}{R^{\sigma_\mb{g}^{\lambda_1}\times\sigma_\mb{h}^{\lambda_2}, \psi}}.
	\label{}
\end{equation}
Now we set $\lambda_1=\lambda_2=+$ and notice $\sigma_\mb{g}^+\times\sigma_\mb{h}^+=\psi^{n(\mb{g},\mb{h})}\times\sigma_{\mb{gh}}^+$, we arrive at
\begin{equation}
	\theta(\mb{g},\mb{h})=(-1)^{n_\mb{g,h}}\frac{R^{\sigma_\mb{g}^+\psi} R^{\sigma_\mb{h}^+\psi}}{R^{\sigma_\mb{gh}^+\psi}}.
	\label{}
\end{equation}
  We can further show that $R^{\sigma_\mb{g}^+\psi}=\pm 1$ by considering the heptagon equation for inverse braiding with the three outgoing lines being $\psi, \psi, \sigma_\mb{g}^\lambda$. This is basically what we need, i.e. $\theta$ is $\mathbb{Z}_2$-cohomologically equivalent to $(-1)^n$.

Therefore, the necessary condition for the extension to exist is that the $4$-cocycle:
\begin{equation}
	O(\mb{g},\mb{h},\mb{k},\mb{l})=(-1)^{n_{\mb{g},\mb{h}}n_{\mb{k},\mb{l}}}
	\label{}
\end{equation}
is in the trivial cohomology class, i.e. $[O]=0$. Once the obstruction vanishes, different solutions of $\nu$ are related to each other by a $3$-cocycle.

\onecolumngrid
\section{Projective Pentagon Equation}
\label{sec:app1}

In this section we give a different derivation of the obstruction mapping, which is closer in spirit to the group super-cohomology theory.

Instead of considering all the defects $\sigma_\mb{g}^\pm$, we take a representative, e.g. $\sigma_\mb{g}^+$ from each $\mb{g}$-sector. To account for the fusion rules, we allow the fusion space to be fermionic, i.e. $\mathbb{Z}_2$-graded. In our case, the fermion parity of the fusion space of $\sigma_\mb{g}$ and $\sigma_\mb{h}$ is completely determined by $\mb{g}$ and $\mb{h}$: it is just $(-1)^{n(\mb{g},\mb{h})}$. More general constructions of fermionic TQFTs have been studied in \Refs{Gu_2010ftop, GuPRB2014}.

With this modification, now the $F$ move becomes an operator possibly connecting states with different fermion parities.

\begin{equation}
\begin{tikzpicture}[baseline={($ (current bounding box) - (0,10pt) $)},scale=0.25]
	\draw[very thick] (0, 8) node[above]{$\mb{g}$} -- (2, 6) ;
	\draw[very thick ]  (2, 6) node[below]{$1$} -- (4, 4) node[below]{$2$};
	\draw[very thick]  (4, 8) node[above]{$\mb{h}$}-- (2, 6);
    \draw[very thick] (4,4) -- (6,2) ;
	\draw[very thick](8, 8) node[above]{$\mb{k}$}-- (4,4);
\end{tikzpicture}
=\nu(\mb{g},\mb{h},\mb{k})f_2^{n(\mb{gh},\mb{k})}f_1^{n(\mb{g},\mb{h})}f_3^{-n(\mb{h},\mb{k})}f_2^{-n(\mb{g},\mb{hk})}
\begin{tikzpicture}[baseline={($ (current bounding box) - (0,10pt) $)},scale=0.25]
\def\shiftx{15};
\def\shifty{0}
\draw[very thick] (\shiftx+0, \shifty+8) node[above]{$\mb{g}$} -- (\shiftx+2, \shifty+6) ;
\draw[very thick]  (\shiftx+2, \shifty+6)  -- (\shiftx+4, \shifty+4) node[below]{$2$};
	\draw[very thick]  (\shiftx+4, \shifty+8) node[above]{$\mb{h}$}-- (\shiftx+6, \shifty+6) node[below]{$3$};
    \draw[very thick] (\shiftx+4,\shifty+4) -- (\shiftx+6,\shifty+2);
	\draw[very thick](\shiftx+8, \shifty+8) node[above]{$\mb{k}$}-- (\shiftx+6, \shifty+6);
	\draw[very thick] (\shiftx+6, \shifty+6) -- (\shiftx+4,\shifty+4);
\end{tikzpicture}.
\end{equation}
Here when we write $f$ we really mean the fermionic creation operator while $f^{-1}$ means the annihilation one.

\begin{figure}
	\includegraphics[width=0.8\textwidth]{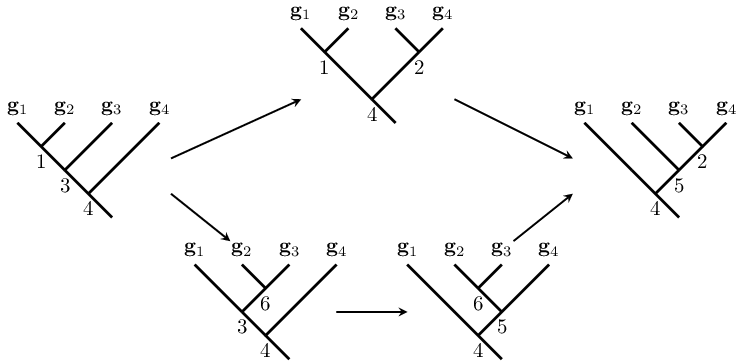}
    \caption{The Pentagon equation enforces the consistency between different sequences of $F$ moves starting and ending with the same fusion trees.}
    \label{fig:pentagon}
\end{figure}

Let us substitute the fermionic $F$ moves into the pentagon equation. We will not explicitly write the $\nu$ factors since they just give the standard $4$-coboundary, and focus on the fermionic part. First evaluate the upper path of the Pentagon equation:
\begin{equation}
\begin{split}
f_4^{n(\mb{g}_1\mb{g}_2\mb{g}_3,\mb{g}_4)}&f_3^{n(\mb{g}_1\mb{g}_2,\mb{g}_3)}f_2^{-n(\mb{g}_3,\mb{g}_4)}f_4^{-n(\mb{g}_1\mb{g}_2,\mb{g}_3\mb{g}_4)}
f_4^{n(\mb{g}_1\mb{g}_2,\mb{g}_3\mb{g}_4)}f_1^{n(\mb{g}_1,\mb{g}_2)}f_5^{-n(\mb{g}_2,\mb{g}_3\mb{g}_4)}f_4^{-n(\mb{g}_1,\mb{g}_2\mb{g}_3\mb{g}_4)}\\
&=f_4^{n(\mb{g}_1\mb{g}_2\mb{g}_3,\mb{g}_4)}f_3^{n(\mb{g}_1\mb{g}_2,\mb{g}_3)}f_2^{-n(\mb{g}_3,\mb{g}_4)}f_1^{n(\mb{g}_1,\mb{g}_2)}f_5^{-n(\mb{g}_2,\mb{g}_3\mb{g}_4)}f_4^{-n(\mb{g}_1,\mb{g}_2\mb{g}_3\mb{g}_4)}.
\end{split}
\end{equation}
Then the lower path gives:
\begin{equation}
\begin{split}
f_3^{n(\mb{g}_1\mb{g}_2,\mb{g}_3)}f_1^{n(\mb{g}_1,\mb{g}_2)}f_6^{-n(\mb{g}_2,\mb{g}_3)}f_3^{-n(\mb{g}_1,\mb{g}_2\mb{g}_3)}\times f_4^{n(\mb{g}_1\mb{g}_2\mb{g}_3,\mb{g}_4)}f_3^{n(\mb{g}_1,\mb{g}_2\mb{g}_3)}f_5^{-n(\mb{g}_2\mb{g}_3,\mb{g}_4)}f_4^{-n(\mb{g}_1,\mb{g}_2\mb{g}_3\mb{g}_4)}\\
\times f_5^{n(\mb{g}_2\mb{g}_3,\mb{g}_4)}f_6^{n(\mb{g}_2,\mb{g}_3)}f_2^{-n(\mb{g}_3,\mb{g}_4)}f_5^{-n(\mb{g}_2,\mb{g}_3\mb{g}_4)}
\end{split}
\label{eqn:low2}
\end{equation}
We move the first $f_4^{s(\mb{g}_1\mb{g}_2\mb{g}_3,\mb{g}_4)}$ to the left most and then $f_4^{-s(\mb{g}_1,\mb{g}_2\mb{g}_3\mb{g}_4)}$ to the right most and the fermionic signs resulting is the following:
\begin{equation}
(-1)^{n(\mb{g}_1\mb{g}_2\mb{g}_3,\mb{g}_4)[n(\mb{g}_1\mb{g}_2,\mb{g}_3)+n(\mb{g}_1,\mb{g}_2)-n(\mb{g}_2,\mb{g}_3)-n(\mb{g}_1,\mb{g}_2\mb{g}_3)]}(-1)^{-n(\mb{g}_1,\mb{g}_2\mb{g}_3\mb{g}_4)[n(\mb{g}_2\mb{g}_3,\mb{g}_4)+n(\mb{g}_2,\mb{g}_3)-n(\mb{g}_3,\mb{g}_4)-n(\mb{g}_2,\mb{g}_3\mb{g}_4)]}
\label{eqn:sign1}
\end{equation}
From the $2$-cocyle condition
\begin{equation}
\omega(\mb{g}_1,\mb{g}_2\mb{g}_3)\omega(\mb{g}_2,\mb{g}_3)=\omega(\mb{g}_1\mb{g}_2,\mb{g}_3)\omega(\mb{g}_1,\mb{g}_2),
\end{equation}
together with $\omega=(-1)^n$, we have
\begin{equation}
n(\mb{g}_1,\mb{g}_2\mb{g}_3)+n(\mb{g}_2,\mb{g}_3)=n(\mb{g}_1\mb{g}_2,\mb{g}_3)+n(\mb{g}_1,\mb{g}_2)\,(\text{mod } 2).
\end{equation}
Therefore the term \eqref{eqn:sign1} is identically $1$. 

After this manipulation the expression \eqref{eqn:low2} can be greatly simplified:
\begin{equation}
f_4^{n(\mb{g}_1\mb{g}_2\mb{g}_3,\mb{g}_4)}f_3^{n(\mb{g}_1\mb{g}_2,\mb{g}_3)}f_1^{n(\mb{g}_1,\mb{g}_2)}f_2^{-n(\mb{g}_3,\mb{g}_4)}f_5^{-n(\mb{g}_2,\mb{g}_3\mb{g}_4)}f_4^{-n(\mb{g}_1,\mb{g}_2\mb{g}_3\mb{g}_4)}.
\label{eqn:low}
\end{equation}
So the upper and the lower paths \eqref{eqn:low} just differs by exchanging $f_1^{n(\mb{g}_1,\mb{g}_2)}$ and $f_2^{-n(\mb{g}_3,\mb{g}_4)}$ which results in a sign:
\begin{equation}
O(\mb{g}_1,\mb{g}_2,\mb{g}_3,\mb{g}_4)=(-1)^{n(\mb{g}_1,\mb{g}_2)n(\mb{g}_3,\mb{g}_4)}.
\end{equation}
In order to satisfy the pentagon equation, this sign has to be compensated by $\di\nu$:
\begin{equation}
	(\di\nu)(\mb{g},\mb{h},\mb{k},\mb{l})=(-1)^{n(\mb{g},\mb{h})n(\mb{k},\mb{l})}.
	\label{}
\end{equation}
This is the obstruction-free condition.

\twocolumngrid

\bibliography{./spt}

\end{document}